\documentclass[a4paper,10pt]{article}
\usepackage{cite}
\usepackage[margin=1in]{geometry}

\usepackage{amsmath,amssymb,amsfonts}
\usepackage{algorithmic}
\usepackage{graphicx}
\usepackage{textcomp}
\usepackage{wrapfig,colortbl}

\usepackage{fancyhdr}
\definecolor{abstractbg}{rgb}{1,0.969,0.914}
\def\BibTeX{{\rm B\kern-.05em{\sc i\kern-.025em b}\kern-.08em
    T\kern-.1667em\lower.7ex\hbox{E}\kern-.125emX}}

\begin{document}
\title{A comprehensive equivalent circuit model for high overtone bulk acoustic resonators (HBARs)}
\author{Vikrant J. Gokhale, and Brian P. Downey.
\thanks{This work was supported by the Office of Naval Research. }
\thanks{Authors are with the Electronic Science and Technology Division of the US Naval Research Laboratory (NRL), Washington DC, USA (e-mail: vikrant.j.gokhale.civ@us.navy.mil). }
%
}
\maketitle
\begin{abstract}
This paper presents a new and comprehensive equivalent circuit model for high overtone bulk acoustic resonators (HBARs). HBARs demonstrate several very sharp resonance modes distributed nearly periodically over a very wide frequency range. This spectrum response of HBARs offers unique advantages but poses significant modeling challenges. The proposed circuit incorporates and models the unique physical components of the HBAR: piezoelectric transducer, substrate (a perfectly periodic multimode cavity), piezoelectric coupling, and critically, the imperfectly matched transducer-substrate interface which imparts characteristic aperiodicity to the HBAR mode spectrum. By judicious use of fixed, periodic, or tightly constrained virtual lumped-element branches, and sets of branches, the model retains clear and intuitive links to the physical device, while reducing the complexity needed for fitting dense, broadband datasets. We demonstrate the validity and power of this model by simultaneously fitting measured data for 61 modes of a GaN/NbN/sapphire HBAR over a span of 1 GHz, and extracting modal parameters such as quality factors and coupling coefficients. We show that this new model is compact and yet scalable: by leveraging the inherent internal relationships in an HBAR, the model can be easily expanded to include multiple transducer overtones and envelopes, multiple distinct transducers, and spurious modes. In addition to fitting measured datasets, the new model can also be used to easily analyze various perturbations to the nominal state of the HBAR. We expect the new model to be useful for the design of classical HBAR-based oscillators, filters, and sensors, and for the integration of HBARs into quantum circuits.
\end{abstract}

%

\section{Introduction}
\label{sec:introduction}
\label{sec:SecI}

High overtone bulk acoustic resonators (HBARs) are solidly-mounted micromechanical cavity resonators with unique attributes that set them apart from other bulk or surface wave resonators. The structure, operational principles, and applications of HBARs have been described at length in the literature and will only be briefly discussed here for context \cite{RN2380,RN1832}. Canonically, HBARs are composite devices comprised of a thin parallel plate metal-piezoelectric-metal transducer in mechanical contact with a much thicker substrate. For the purposes of this discussion, we consider the metal-piezoelectric-metal stack to be a single entity; the transducer. On application of an input drive signal, the piezoelectric transducer generates acoustic waves and transmits them into the substrate, which acts as an efficient cavity for the fundamental mechanical resonance frequency, and all subsequent harmonic modes (overtones) that can be sustained. Thus, the primary distinguishing characteristic of the HBAR is its massively multi-mode frequency spectrum, often containing a very large ensemble of sharp resonance modes spaced nearly periodically. A similar laterally-oriented device, the lateral overtone bulk acoustic resonator (LOBAR) \cite{RN2012,RN2024}, can be considered an HBAR variant for the purposes of this discussion.

Mason’s acoustic transmission line model and the subsequent expansion by Cheeke give us a complete analytical model for the electrical input impedance of HBARs for known device structures \cite{RN1832}. These models provide deep insight into the physics and behavior of HBARs; however, they require precise information about all relevant material properties for all the constituent layers. Precise information is not often available for thin films, high frequencies (small wavelengths), or across temperatures. This makes the use of these models for predictive design, modeling or post-measurement data fitting challenging for all but the simplest cases.


Another, and arguably more practical, approach towards model fitting and parameter extraction of mechanical resonators is the lumped-element equivalent circuit approach, where key parameters of a vibrational mode of a mechanical resonator with one degree of freedom (stiffness $K$, mass $M$, damping $B$) are mapped onto equivalent parameters (capacitance $C$, inductance $L$, resistance $R$) of a virtual electrical resonator comprised of discrete components. This follows from the fundamental equivalence between damped harmonic oscillators in the mechanical and electrical domains, such that: 
\begin{equation}F=M\ddot{x}+B\dot{x}+Kx\label{eq1}\end{equation}
\begin{center}
	\textdownarrow\textdownarrow \qquad \qquad \textuparrow\textuparrow
\end{center}
\begin{equation}V=L\ddot{q}+R\dot{q}+C^{-1}q\label{eq2}\end{equation}

Here \(F\) and \(x\) are force and displacement in the mechanical domain, \(V\) and \(q\) are voltage and charge in the electrical domain, and the single- and double-dot notation denotes the first and second derivatives of the variable with respect to time (Newton's notation). The virtual lumped elements \(L\), \(C\), and \(R\)  correspond to the physical quantities modal mass (\(M\)), modal compliance (or inverse of stiffness, \(K^{-1}\)), and modal damping (\(B\)) respectively. The relationships between various elements give us insight into the resonance frequency ($\omega$), the quality factor ($Q$), and the coupling coefficient ($k^2$) of the resonator.

\begin{figure}
	\centering
	\includegraphics[width=0.7\linewidth]{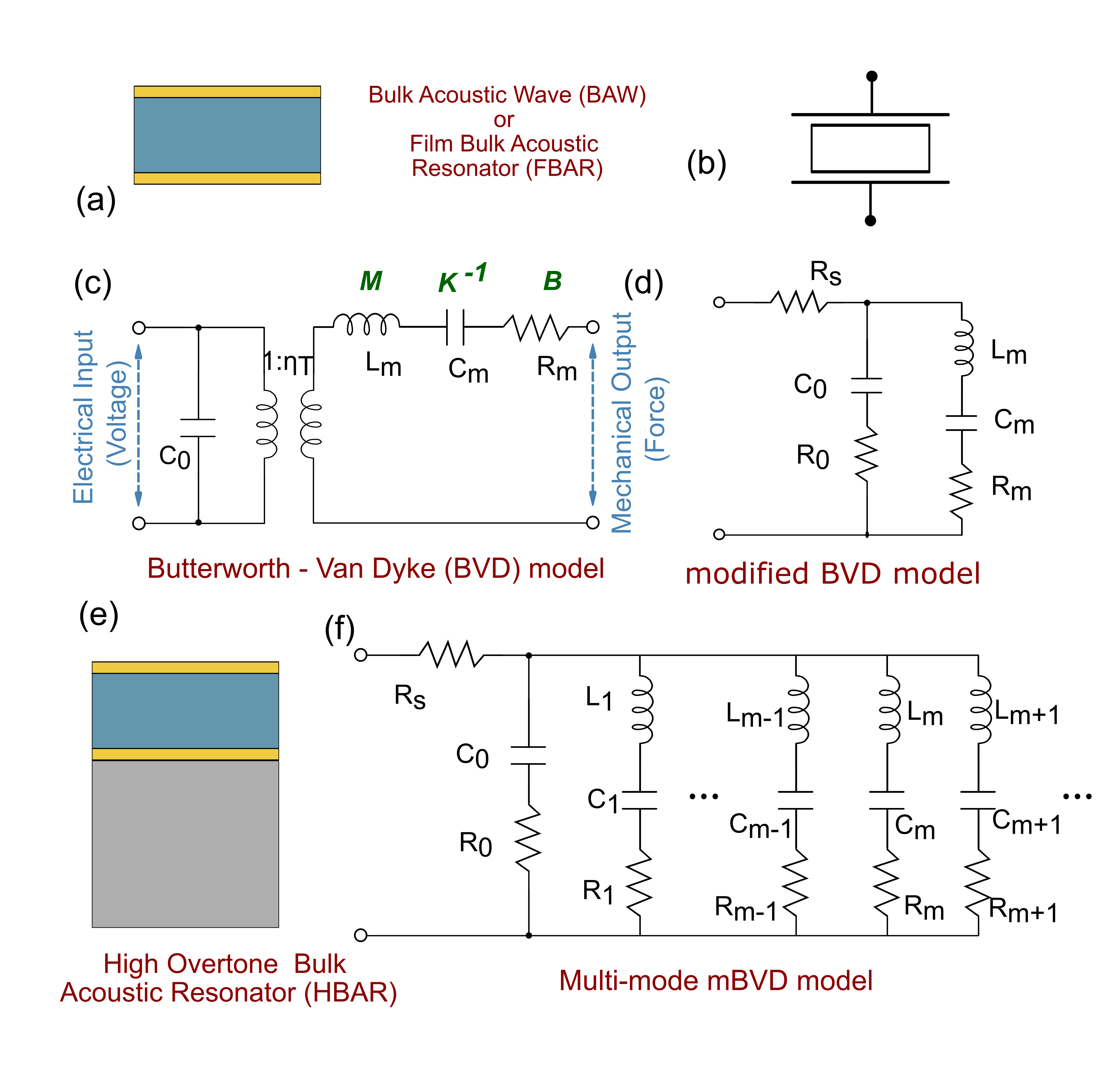}
	\caption{The evolution of equivalent circuit models used to describe single and multi-mode resonators. (a) The simple quartz crystal resonator, BAW, or FBAR along with (b) its basic circuit symbol. The (c) BVD, and (d) the mBVD models have been used to model the mechanical resonance of these crystals. (d) The HBAR is a piezoelectric transducer mounted on a substrate, and can be represented by an extension of the mBVD model: (e) the multi-mode BVD model where each branch represents one mode. This is a simplistic model that obscures the links to the physical HBAR and introduces an arbitrarily large set of parameters for data fitting and parameter extraction. This paper presents a comprehensive and yet more intuitive equivalent circuit model for HBARs.}
	\label{fig:fig1}
\end{figure}

This equivalent electrical circuit approach, the Butterworth-van Dyke (BVD) network, was developed over a hundred years ago and has proved invaluable in describing the electrically-driven vibrations of quartz crystal plates at mechanical resonance \cite{RN3015,RN3008}. A modified Butterworth-van Dyke (mBVD) model was proposed more recently by Larson \textit{et al} \cite{RN2311} to account for electrical loss in thin film piezoelectrics (Fig. \ref{fig:fig1}(a-c)). These models have been immensely successful at modeling and predicting the behavior of quartz crystal microbalances (QCMs), quartz oscillators, and film bulk acoustic resonators (FBARs). The mBVD model was subsequently generalized to describe a variety of MEMS resonators and acoustically coupled filters, including flexural mode resonators, surface acoustic wave resonators, extensional, shear, and Lamé mode resonators, resonators with one or two ports, resonators subject to mass loading, and involving various forms of damping \cite{RN2995,RN2018,RN2011,RN3017}.

The mBVD model has been used to model multiple modes and overtones of FBARs and other resonators \cite{RN3020}. The authors too have recently used a similar phenomenological ‘multi-mode mBVD’ approach (Fig. \ref{fig:fig1}(d-e)) to model HBAR modes over a limited range of frequencies, simply by adding one virtual \(LCR\) branch to account for each HBAR mode and tuning the parameters of each branch individually \cite{RN2173,RN2249}. 

However, even a single HBAR can cover a very wide frequency range, with hundreds or even thousands of very sharp HBAR modes. The multi-mode BVD and mBVD models are infinitely extensible in theory, but using them to fit measured data over hundreds of modes quickly becomes a brute-force numerical exercise. While advanced computational techniques are quite capable of handling large datasets, the biggest drawback to this computational-first approach to modeling HBARs is that it obscures the clear physical analogies and insights provided by the mBVD model. 

The goals of this paper are to present a new lumped-element HBAR equivalent circuit model which 1) can model the unique spectral characteristics of HBARs in a compact manner, 2) retain a meaningful mapping of each virtual network element to a parameter in the physical domain, 3) accommodate a variety of design configurations or model perturbations, and 4) provide a simple kernel for fitting large measured data sets using advanced computational techniques including physics-informed machine learning (PIML) routines. 

We divide the rest of the paper into four major sections. In Section \ref{sec:SecII} , we build the new HBAR equivalent circuit. Section \ref{sec:SecIII} lays out a simple algorithm for data fitting and parameter extraction and uses measured HBAR data to validate both the model and fitting algorithm. Section \ref{sec:SecIV} extends the basic HBAR model by adding higher order transducer envelopes or multiple transducers on the same substrate while retaining its compact form. Finally, Section \ref{sec:SecV} discusses the use of this new equivalent circuit to model a variety of dispersive and dissipative perturbations to the HBAR, highlighting the model's utility for the design and analysis of HBAR-based applications.

\section{The HBAR Equivalent Circuit Model}
\label{sec:SecII}
In this section, we start with a brief review of the structural and functional components of HBARs, and proceed to build the basic equivalent circuit model.

\subsection{Building the basic model}
Structurally, the HBAR is made of two distinct sub-components: a thin metal-piezoelectric-metal transducer mounted on a much thicker substrate. Functionally, it is more useful to think of the basic HBAR device as three separate elements linked sequentially by two coupling mechanisms. 
The first element is purely electrical; the driving input RF signal is applied across the static capacitance of the transducer. This non-ideal capacitor is described by the pure dielectric response of the piezoelectric transducer $C_0 = A\epsilon_0\epsilon_r/t $, and \(R_0\) which represents the  purely electrical losses in the piezoelectric film. An additional input series resistance \(R_s\) lumps all resistive losses external to the dielectric, including electrical resistance of the transducer electrodes, and the experimental apparatus (e.g., cable, probe, and contact resistance). 

The second element is the mechanical response of the piezoelectric transducer, which is modeled as a virtual branch containing the series combination of \(L_T\), \(C_T\), and \(R_T\). These two elements are coupled by piezoelectric transduction, which converts power from the electrical domain to the mechanical domain. This coupling is modeled as a virtual transformer with a turn ratio of 1:\(\eta_T\) where \(\eta_T\)  is the power transduction efficiency from the electrical domain to the mechanical domain, or vice-versa \cite{RN3017}. The resonance frequency of the piezoelectric transducer (\(\omega_T\)), the intrinsic (or unloaded) mechanical quality factor (\(Q_T\)), and the piezoelectric coupling coefficient (\(k_T^2 \))  are given below.

\begin{equation} \omega_T = \frac{1}{\sqrt{L_T C_T}}     \label{eq3} \end{equation}

\begin{equation} Q_T = \frac{\omega_T L_T}{R_T} = \frac{1}{\omega_T R_T C_T}    \label{eq4} \end{equation}

\begin{equation} k_T^2 = \frac{\pi^2}{8} \frac{C_T}{C_0} \left(\frac{C_0 - C_T}{C_0} \right)     \label{eq5} \end{equation}

Note that all mechanical damping is lumped into the virtual dissipative element \(R_T\). Up to this point, we have followed the mBVD model of the FBAR exactly, and all elements so far have their usual meaning and significance.

The third element of the HBAR equivalent circuit must represent the substrate and is the critical point of departure from the conventional BVD or mBVD models. For this argument, we assume that the substrate is made of a single material with a constant longitudinal acoustic velocity \(v\), is semi-infinite in plane, and has a finite thickness  \(t_S\). The substrate acts as an acoustic cavity for the acoustic frequency \(\omega_1\) and wavelength \(\lambda_1\) that meet the condition \(\lambda_1 = 2t_S \), and all its higher harmonics. Thus, the substrate can efficiently confine all \(m\) cavity phonon modes that meet the conditions: 

\begin{equation} \lambda_m =  \frac{\lambda_1}{m}= \frac{2t_S}{m}   \label{eq6} \end{equation}
\begin{equation} f_m = \frac{\omega_m}{2\pi} = m\left(\frac{\omega_1}{2\pi}\right) = \frac{v}{\lambda_m} =m\left(\frac{v}{2t_S}\right)  \label{eq7} \end{equation}

This represents a perfectly periodic acoustic or phononic cavity. For our HBAR equivalent circuit, we model this multi-mode cavity mode as \(m\) parallel branches, each comprised of the series combination of \(L_m\), \(C_m\), and \(R_m\), such that
\begin{equation} \omega_m =  \frac{1}{\sqrt{L_m C_m}} = \frac{m}{\sqrt{L_1 C_1}}   \label{eq8} \end{equation}
\begin{equation} L_m = \frac{L_1}{m} \quad;\quad C_m = \frac{C_1}{m} \label{eq9}\end{equation}

The physical analogy of (\ref{eq1}) and (\ref{eq2}) between the electrical and mechanical oscillators remains true for higher modes; we can easily write the non-dissipative virtual elements \(L_m\) and \(C_m\) as well as the modal mass \(M_m\), and modal stiffness \(K_m\) as functions of \(m\): 

\begin{equation} \omega_m =  {\sqrt{\frac{K_m}{M_m}}} =  m{\sqrt{\frac{K_1}{M_1}}}   \label{eq10} \end{equation}
\begin{equation} M_m = \frac{M_1}{m} \quad;\quad K_m = K_1 m \label{eq11}\end{equation}

Note that the modal mass \(M_m\) decreases as a function of \(m\), while modal stiffness \(K_m\) increases with \(m\). 
The analogy is often made that the substrate behaves like an acoustic Fabry-Perot cavity. For the bare, homogenous, semi-infinite substrate, this is exactly true, and we should expect the modes to be perfectly periodic, mathematically represented by a Dirac frequency comb. In this ideal situation, the mode spacing or the free spectral range \(FSR\) is constant, and is equal to the frequency of the fundamental mode \(\omega_1\).  

\begin{equation} FSR = \omega_{m+1} - \omega_m = \omega_1	\label{eq12}\end{equation}

\begin{figure*}
	\centering
	\includegraphics[width=1\linewidth]{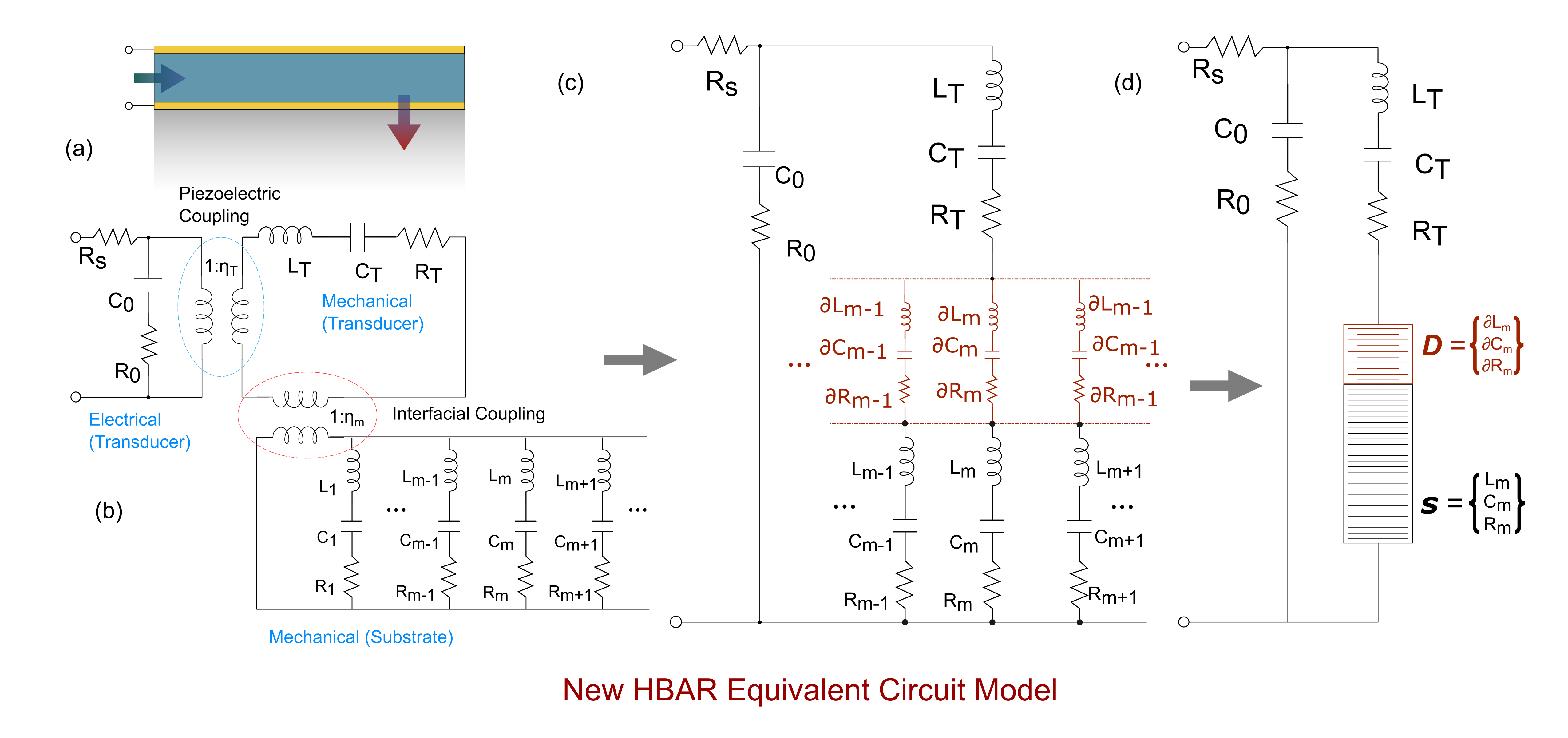}
	\caption{The (a) canonical HBAR can be represented as (b) an equivalent circuit model with three elements (the purely electrical branch and the mechanical branch of the transducer, and the purely mechanical, strictly periodic branches of the substrate) and two internal coupling mechanisms (piezoelectric and interfacial) which are modeled as virtual transformers. (c) A discretized version of this circuit retains the strict periodicity and internal relationships between the substrate branches and introduces detuning couplers for each mode that account for any aperiodicity. (d) A compact representation of the HBAR equivalent circuit.}
	\label{fig:fig2}
\end{figure*}

In a practical HBAR device, the substrate is asymmetrically loaded by the piezoelectric transducer with a finite thickness \(t_T\), where \(t_T \ll t_S \). This introduces two non-idealities to the perfect bare substrate: 1) the total cavity length is now the sum of the thicknesses of the substrate and transducer such that, \(t_{total}=t_S+t_T \), and 2) this composite cavity no longer has homogenous acoustic properties. An acoustic impedance mismatch between the transducer and the substrate will result in imperfect transmission of acoustic power across the interface. This is a well understood phenomenon; several analyses in the literature explain it, model it analytically, and verify it experimentally \cite{RN2199,RN2249,RN3019,RN2732}. In terms of lumped elements, the characteristic acoustic impedances of the transducer \(Z_T\) and substrate \(Z_m\) can be written as functions of their respective mass and stiffness, or their equivalent virtual electrical analogues: 

\begin{equation} Z_T = \sqrt{K_T M_T} =  \sqrt{L_T/C_T}	\label{eq13}\end{equation}
\begin{equation} Z_m = \sqrt{K_m M_m} =  \sqrt{L_m/C_m}	\label{eq14}\end{equation}

For most HBARs \(Z_T \neq Z_m\), i.e., there is acoustic impedance mismatch at the interface between transducer and substrate. In addition to imperfect power transmission, the inhomogeneous structure of the HBAR cavity and the acoustic impedance mismatch at the transducer-substrate interface also result in wavelength and frequency dispersion. In the HBAR modal spectrum, this dispersion manifests as a mode-dependent, discrete set of frequency shifts for all $m$ cavity phonon modes, appearing as a small modulation, ‘ripple’, or aperiodicity in the mode spacing spectrum. This modulation effectively breaks the perfect periodicity of the ideal substrate. 

To account for these small frequency shifts in the new equivalent circuit model, we define the fractional aperiodicity \(a_m\)  as the small fractional deviation from perfect periodicity such that the exact frequency of the loaded HBAR mode m is given by

\begin{equation} \omega_m^{'} = \omega_m + a_m \omega_m 		\label{eq15}\end{equation}
\begin{equation} a_m = \frac{\omega_m^{'}-\omega_m}{\omega_m} 	\label{eq16}\end{equation}

The fractional aperiodicity \(a_m\) can take positive or negative values, and \(|a_m| \ll 1\) . Exact values for \(a_m\)  can be analytically calculated for every mode across the HBAR spectrum using the analytical transmission line model (Cheeke model) for known material properties, or easily extracted from measured HBAR spectra. In HBARs with low acoustic impedance mismatch, values of \(a_m\)  are on the order of \(\pm 10^{-3}\) \cite{RN2199,RN2249,RN3019,RN2732}. 

The new HBAR equivalent circuit model must account for this dispersive power transfer from transducer to the ideal substrate and the accompanying aperiodicity. We model this imperfect coupling as a virtual interface transformer, with a turn ratio of 1: \(\eta_m\), where \(\eta_m\) denotes the power transfer efficiency across the interface for each mode m. Unlike the piezoelectric transformer, the virtual interface transformer converts purely to and from the mechanical domain.

\subsection{New circuit elements and the complete HBAR model}

Based on the discussion above, we can now recast the HBAR equivalent circuit model (with its three elements and two coupling transformers) into branches or sets of branches. The purely electrical branch, including the series resistance is denoted by $ \textbf{\textit{E}}=\{C_0, R_0, R_S\} $ and the virtual branch representing the transducer is denoted by $ \textbf{\textit{T}}=\{L_T, C_T, R_T\} $. The virtual branches representing the perfectly periodic substrate are combined into a set $ \textbf{\textit{S}}=\{L_m, C_m, R_m\} $, for all integer values of \(m\). We discretize the interface transformer by introducing a new set of virtual branches $ \textbf{\textit{D}}=\{\delta L_m, \delta C_m, \delta R_m\} $. Each \(m^{th}\) branch of $ \textbf{\textit{D}} $ is in series with each corresponding \(m^{th}\) branch of $ \textbf{\textit{S}} $. The non-dissipative elements $\delta L_m$ and $\delta C_m$ correspond to the differential density and stiffness across the interface for each mode, while the dissipative elements $\delta R_m$ represent wavelength-dependent scattering losses at the interface in a practical device with non-zero roughness \cite{RN1711}. In effect, $ \textbf{\textit{D}} $ is the set of detuning couplers that connects the transducer branch $ \textbf{\textit{T}} $ with individual substrate branches in $ \textbf{\textit{S}} $ and impart aperiodicity to the overall network. 

From a circuit theory or numerical fitting perspective, the non-dissipative coupling element linking $ \textbf{\textit{T}} $ and $ \textbf{\textit{S}} $ could have been either purely inductive or capacitive. However, the aperiodicity is a function of acoustic impedance mismatch at the interface, and consequently should include both inductive and capacitive elements to better mirror physical reality. 

Figure \ref{fig:fig2} shows the evolution of the new HBAR equivalent circuit model which retains clear links between physical elements, coupling mechanisms, and the lumped elements and element sets that can be used to describe the full behavior of the HBAR. For completeness, we can  formally write the HBAR equivalent circuit as a virtual network that includes the superset $ \textbf{\textit{H = \{E, S, T, D}\}} $. The modal equations for any mode \(m\) of the HBAR can now be written as:

\begin{equation} \omega_m^{'} = \frac{1}{\sqrt{L_m^{'} C_m^{'}}}     \label{eq17} \end{equation}
\begin{equation} L_m^{'} = L_T + \delta L_m + L_m     \label{eq18} \end{equation}
\begin{equation} \frac{1}{C_m^{'}} = \frac{1}{C_T} + \frac{1}{\delta C_m} + \frac{1}{C_m} \label{eq19} \end{equation}
\begin{equation} R_m^{'} = R_S + R_T + \delta R_m + R_m     \label{eq20} \end{equation}

We reemphasize that $ \textbf{\textit{D}} $ represents the infinitesimal interface between two components, and not a physical layer by itself. Numerically, $ \textbf{\textit{D}} $ should have a small impact on the effective mode equations, i.e., $\delta L_m \ll L_m $, $\delta C_m \gg C_m $, and $\delta R_m \ll R_m $. For a well-designed HBAR, for specific applications where we only operate over a small range of \(m\), $ \textbf{\textit{D}} $ could be considered negligible and simply omitted if we want to ignore small levels of aperiodicity.  

Defining $ \textbf{\textit{S}} $ and $ \textbf{\textit{D}} $ separately gives us the advantage of retaining the perfect periodicity and internal harmonic relationship between the elements of $ \textbf{\textit{S}} $, while modeling the aperiodicity with $ \textbf{\textit{D}} $. Defining $ \textbf{\textit{D}} $ with inductive, capacitive, and resistive elements retains the direct links between the physics of the HBAR and the equivalent circuit. Both sets are valid over all integer values of  \(m\), and thus can be used to describe higher order transducer envelopes without requiring more circuit elements (see Section \ref{sec:SecIV} A).

\subsection{Modal Coupling Coefficients and Quality Factors}
The effective coupling coefficient for each HBAR mode \((k_m^2 )\) describes the relationship between the input electrical energy and the mechanical energy in the substrate, coupled via the piezoelectric transducer and the interface between the transducer and substrate, and is given by: 

\begin{equation} k_m^2 = \frac{\pi}{8} \frac{C_m^{'}}{C_0} \left( \frac{C_0-C_m^{'}}{C_0} \right)   \label{eq21} \end{equation}

Combining equations (\ref{eq5}), (\ref{eq9}), and (\ref{eq19}), we can now write $k_m^2$  across the spectrum as a function of the mode number m, the fixed elements \(C_0\), \(C_T\), \(C_1\), and a single freely variable element \(\delta C_m\). This also allows us to easily describe the relationship between the piezoelectric coupling coefficient $k_T^2$ and $k_m^2$. Note that in most HBARs with substrates that are much thicker than the transducer, \(m \gg 1 \)and \(k_m^2 \ll k_T^2\). In similar configurations such as LOBARs and thin film piezoelectric on substrate (TPoS) resonators we generally deal with smaller values of $m$, resulting in stronger modal coupling coefficients. 

The total modal quality factors for the HBAR modes are given by:
 \begin{equation} Q_m^{'} = \frac{\omega_m^{'} L_m^{'}}{R_m^{'}} =   \frac{1}{\omega_m^{'} R_m^{'} C_m^{'}}  \label{eq22} \end{equation}

Here, \(R_m^{'}\) includes the sum of all losses in the transducer, in the substrate, at the interface, and the electrical losses from the series resistance. Physically, these losses depend on material properties, heterostructure, design and operating environment, and include phonon damping, electron damping, thermoelastic damping, radiation loss (anchor loss), diffraction loss, interface/surface roughness scattering, and likely other mechanisms that are material or operating regime dependent. \(R_m^{'}\) remains an important unresolvable quantity in the model; we cannot easily isolate its various components from a single experiment and can often only extract the total value. From the modeling perspective, it is still important to keep \(R_S\), \(R_T\), \(\delta R_m\), and \(R_m\) logically separate; with careful design of experiment, it might be possible to operate in regimes where only one or two of these losses dominate, and differential trends in the materials of the transducer and substrate can allow us to separately model the two elements. 

The new HBAR equivalent circuit model can be used to fit experimentally measured data across a wide range of modes and frequencies. By keeping sets $ \textbf{\textit{E}} $, $ \textbf{\textit{S}} $, $ \textbf{\textit{T}} $, and $ \textbf{\textit{D}} $ separate, the new model has more elements than the multi-mode mBVD model. However, two of these sets ($ \textbf{\textit{E}} $, $ \textbf{\textit{T}} $) are fixed and the non-dissipative elements of S retain their strictly period internal relationship. Thus, the seemingly more complex model $ \textbf{\textit{H}} $ reduces to an unknown set $ \textbf{\textit{D}} $ and the unknown dissipative elements \(R_m^{'}\). Initial estimates for the aperiodicity described by $ \textbf{\textit{D}} $ can be calculated by well-known acoustic impedance relationships \cite{RN1832}, or easily estimated from measured data. The only truly free variable is the total loss denoted by the chain of dissipative elements \(R_m^{'}\). This reduction in complexity is a great advantage when the model is used to fit measured data and extract parameters.

\section{Data Fitting and Parameter Extraction}
\label{sec:SecIII}

In this section we present a simple algorithmic approach to fitting measured HBAR data to the model described above and use a representative measured dataset as preliminary validation. 

\subsection{Data fitting algorithm}

\begin{figure*}
	\centering
	\includegraphics[width=0.9\linewidth]{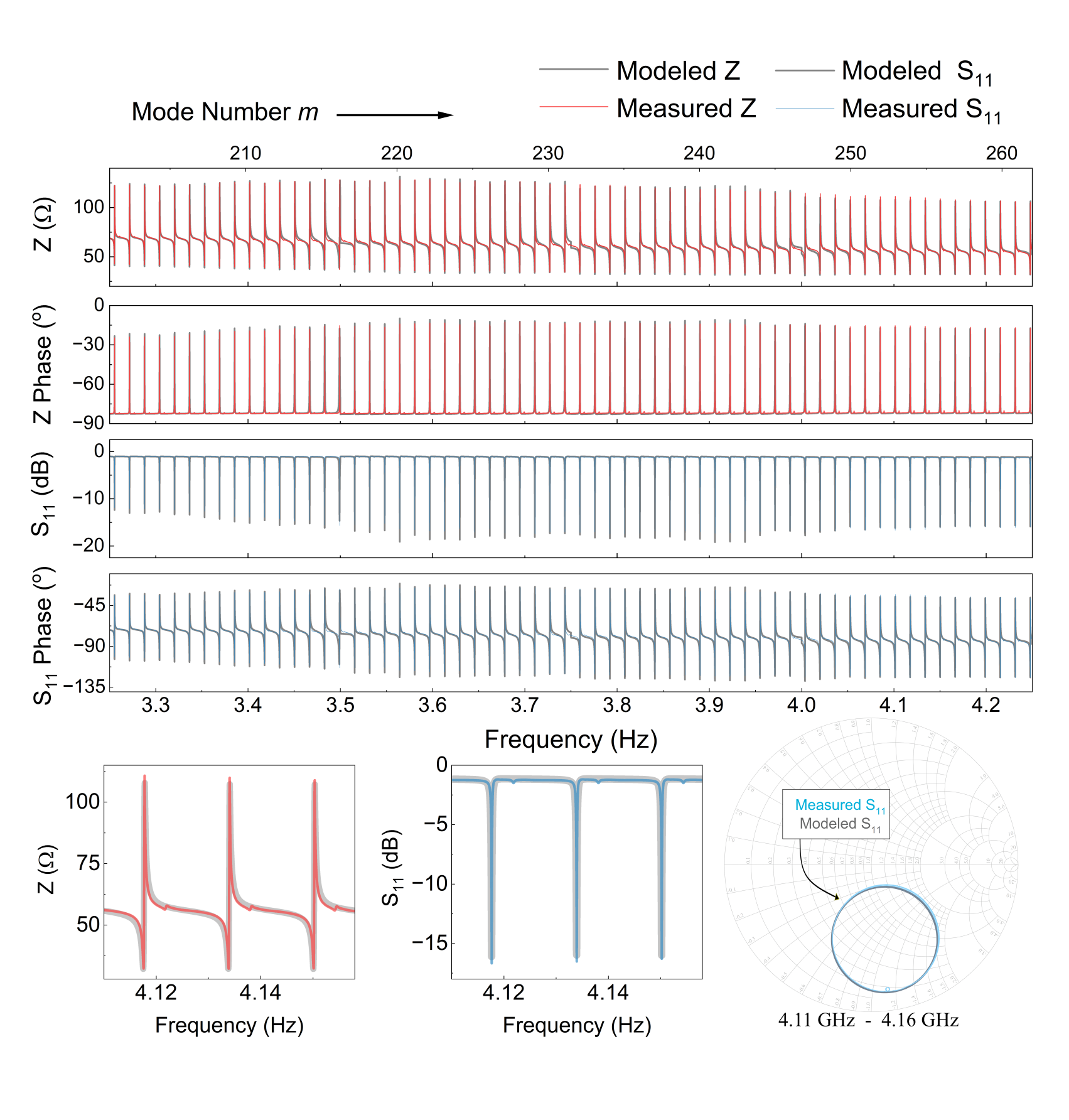}
	\caption{Measured dataset from 3.25 GHz to 4.25 GHz (1 GHz span, 61 HBAR modes) for a GaN/NbN/sapphire HBAR measured at room temperature. The new HBAR equivalent circuit model is used to fit measured data, validating both the model and the fitting algorithm presented here. For clarity, we show a magnified region (arbitrarily chosen) with $|\textbf{Z}|$ and $|\textbf{S}_{11}|$, and a Smith chart representation to demonstrate the close agreement between the model and measured data. Note that we do not attempt to fit small damped pad modes seen in the magnified response;  see Section \ref{sec:SecIV} for details}
	\label{fig:fig3}
\end{figure*}

The algorithm below can be implemented using any circuit simulator capable of modeling RF scattering parameters or impedance. We start with the measured RF reflection spectrum \(\textbf{S}_{11} \) for a single port HBAR spanning multiple HBAR modes. Equivalently, either impedance \((\textbf{Z})\) or admittance \((\textbf{Y})\) parameters can be used. Note that these are all complex quantities, and that both amplitude and phase for all quantities must fit simultaneously and self-consistently.

\begin{enumerate}
	\item 	Establish an initial estimate for the fundamental frequency of the substrate cavity, $f_1= \omega_1 / 2 \pi  =  v ⁄ 2t_S$ based on the thickness and acoustic properties of the substrate. This is approximately equal to the measured FSR. 
	\item Establish initial estimates for \(L_1\) and \(C_1\) so $ \omega_1 =  1/\sqrt{L_T C_T}$ .
	\item 	Directly measure static capacitance \(C_0\)  and leakage resistor \(R_0\)  at low frequencies (near-DC, below \(\omega_1\)). Alternately, use a calculated estimate for \(C_0\) if the dielectric constant and thickness of the piezoelectric film are well known. Direct measurements are preferred since they will include the effects of all parasitic capacitances. 
	\item Using \(C_0\), fix an initial estimage for \(C_T\) based on the expected material \(k_T^2\). 
	\item Fix \(L_T	\) so that $ \omega_1 =  1/\sqrt{L_T C_T}$ is close to the center of the HBAR envelope.
	\item Add S to the network. Use \(L_m = L_1/m\), \(C_m = C_1/m\), and \(\omega_m = m \omega_1\) to get the perfectly periodic HBAR modes, closest to the measured data.
	\item Insert D to model the aperiodicity.
	\item Adjust \(\delta L_m\)  and \(\delta C_m\) to match measured mode frequency. Adjust \(R_s\) and \(R_0\) to fit off-resonance data. Adjust other dissipative parameters in \(R_m^{'}\)  to get a good fit at resonance. 
	\item Fix all parameters except \textbf{\textit{D}} and \(R_m^{'}\). Use an appropriate fitting/optimization routine to fit all required modes.  
	\item Calculate figures of merit \(Q_m^{'}\) and \(k_m^2\) and compare with measured values. 
	
\end{enumerate}

The algorithm shown above is a simple approach aimed at validating the model; more sophisticated fitting can be carried out using physics informed machine learning (PIML) techniques. The advantage of using this new equivalent model is that we transform a fitting and optimization problem involving many independent parameters (as with the multi-mode mBVD model) to an optimization problem with several fixed or strictly mode number dependent quantities along with fewer, narrowly-constrained independent quantities to model the aperiodicity. Further simplifications to this algorithm are possible, leveraging known values of standard components. For example, electrical elements \(R_s\)  , \(R_0\) and \(C_0\), and virtual mechanical elements \(L_1\)  and \(C_1\) will be known or can be measured with high precision in a production facility.

\subsection{A representative parameter extraction example}

As a demonstration, we use the new HBAR equivalent circuit model to fit measured data from a GaN/NbN/sapphire HBAR (800 nm/50 nm/350 $\mu$m). Detailed descriptions of the HBAR will be presented elsewhere; here we present a small subset of the measured data and use it to validate the HBAR model and fitting procedure. A rigorous analysis of parameter sensitivity and algorithmic efficiency for model fitting and optimization are beyond the scope of this paper. 

We used Keysight Advanced Design System (ADS 2025) for modeling HBAR measurements between 3.25 GHz and 4.25 GHz (61 modes) using the procedure in Section III.A. Initial estimates for electrical and transducer parameters are given in Table \ref{table:tableI}. The independent variables \(\delta L_m\), \(\delta C_m\), and \(R_m^{'}\) are optimized using a built-in gradient optimization routine with a goal of matching complex impedance \(\textbf{Z}\). As seen in Fig. \ref{fig:fig3}, the modeled \(\textbf{Z}\) and \(\textbf{S}_{11}\) parameters simultaneously match with the measured data, thus validating the HBAR equivalent model and data-fitting algorithm.

\begin{table}
	\centering
	\caption{Parameter values used for fitting measured HBAR data}
	\label{table:tableI}
	\setlength{\tabcolsep}{3pt}
	\renewcommand{\arraystretch}{1.5}
	\begin{tabular}{p{0.125\textwidth}p{0.125\textwidth}p{0.125\textwidth}}
	
		\hline\hline
		\textbf{Parameter} & \textbf{Initial Estimate} & \textbf{Final Value} \\
		\hline 
		\(Z_0\)& 50 $\Omega$ & 50 $\Omega$  \\ [1ex]
		\hline 
		\(R_s\) & 5 $\Omega$ & 6 $\Omega$ \\ [1ex]
		\hline
		\(C_0\) & 700 fF & 690 fF \\ [1ex]
		\hline
		\(R_0\) & 3.0 $\Omega$ & 2.5 $\Omega$ \\ [1ex]
	
		\hline		\hline
		\(\omega _T /2 \pi\) & 4.00 GHz & 3.92 GHz \\ [1ex]
		\hline
		\(L_T\) & 132 nH & 135 nH \\ [1ex]
		\hline
		\(C_T\) & 12.0 fF & 12.2 fF \\ [1ex]
		
		\hline		\hline
		\(\omega _1 /2 \pi\) & 16.24 MHz & 16.19 MHz \\ [1ex]
		\hline
		\(L_1\) & 9.6 mH & 6.9 mH \\ [1ex]
		\hline
		\(C_1\) & 10 fF & 14 fF \\ [1ex]
		\hline
		\(k_T^2\) & 2.07 \% & 2.14\% \\ [1ex]
		\hline		\hline
	
		\end{tabular}
	\label{tab1}
\end{table}

\begin{figure}
	\centering
	\includegraphics[width=0.7\linewidth]{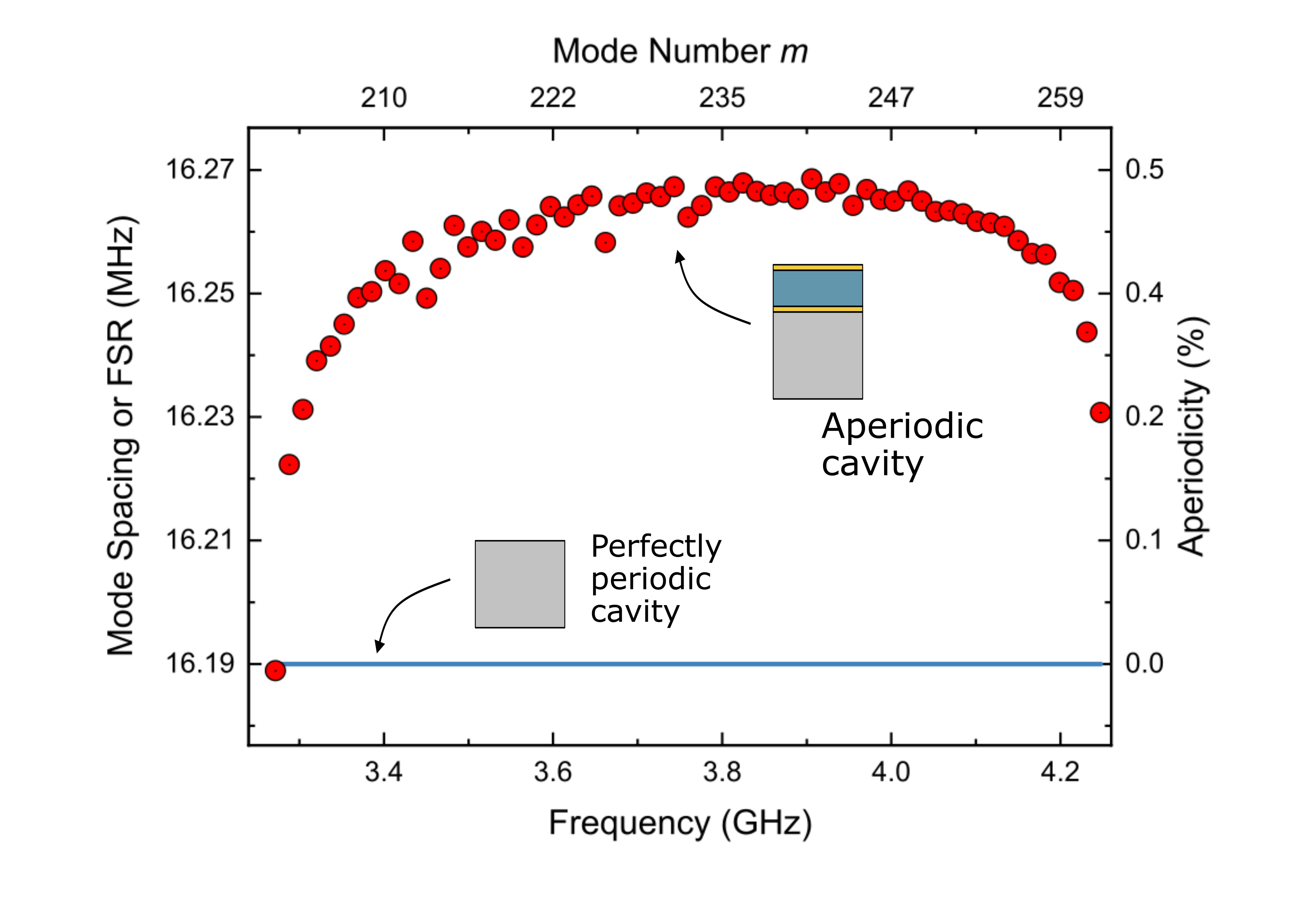}
	\caption[FSR]{Modeled free spectral range (FSR) or the mode spacing captures the aperiodicity associated with the HBAR. The FSR of the perfectly periodic cavity (the semi-infinite sapphire substrate without the transducer) is also shown. As expected, for a GaN/NbN/sapphire heterostructure with low acoustic impedance mismatch, the aperiodicity was less than 0.5 \% . }
	\label{fig:fig4}
\end{figure}

The initial estimate of constant FSR is based on the perfectly periodic cavity, and is only used as a starting point for the model fit. After model optimization and fitting, the final values for the FSR are shown in Fig. \ref{fig:fig4}. The aperiodicity is less than 0.5\% within this spectral range, thus validating \(|a_m | \ll 1 \) and confirming good acoustic impedance matching for the GaN/NbN/sapphire heterostructure.

Using (\ref{eq21}), we can plot the modeled values of $k_m^2$ as a function of $m$ and $k_T^2$. For simple HBARs operating in the first transducer envelope mode, we can simply write this relation as:

\begin{equation}	k_m^2 = \frac{k_T^2}{m}		\label{eq23} \end{equation}

We can also directly calculate \(k_m^2\) from measured series and parallel frequencies \(\omega_s\)  and \(\omega_p\) \cite{RN2311} using the relation: 

\begin{equation}	k_m^2 = \frac{\pi}{4}   \left[ \frac{\omega_s}{\omega_p}  \left( \frac{\omega_p-\omega_s}{\omega_p} \right) \right]  \biggr| _{m} 	\label{eq24} \end{equation}

 Fig.  5 shows both the modeled and directly extracted values of \(k_m^2\), along with their fits to the simple inverse relation in (\ref{eq23}). The inverse linear fits result in values of \(k_T^2\) between 2.16\% and 2.20\%, very close to the initial estimates and expectations for GaN as a piezoelectric transducer \cite{RN2333}. 

A key point highlighted here is that the effective coupling coefficients \(k_m^2\) vary at a rate of \(m^{-1}\) with respect to the substrate dependent HBAR mode or overtone number. For overtones $n$ of the piezoelectric transducer itself, the effective coupling coefficients vary at a rate of \(n^{-2}\) \cite{RN3020,RN1711}. See Section IV.A for a detailed discussion of higher order transducer overtones and envelopes. 

This exercise demonstrates a very important use for the new HBAR equivalent model: to extract $k_T^2$ values for newer piezoelectric materials that are not well characterized. Other techniques have been used to extract unknown values of $k_T^2$ in the past; these often rely on simplifying assumptions about the material heterostructures, or are only valid in specific ranges, or break down at higher frequencies \cite{RN1832,RN361,RN3051} Extraction of $k_T^2$ using the new HBAR equivalent model is more data intensive than simpler analytical methods, but the ability to simultaneously fit hundreds of HBAR modes across a wide frequency range could enable convergence to the values of the unknown $k_T^2$ with high confidence, even under conditions where simple analytical models are not valid. 

\begin{figure}
	\centering
	\includegraphics[width=0.7\linewidth]{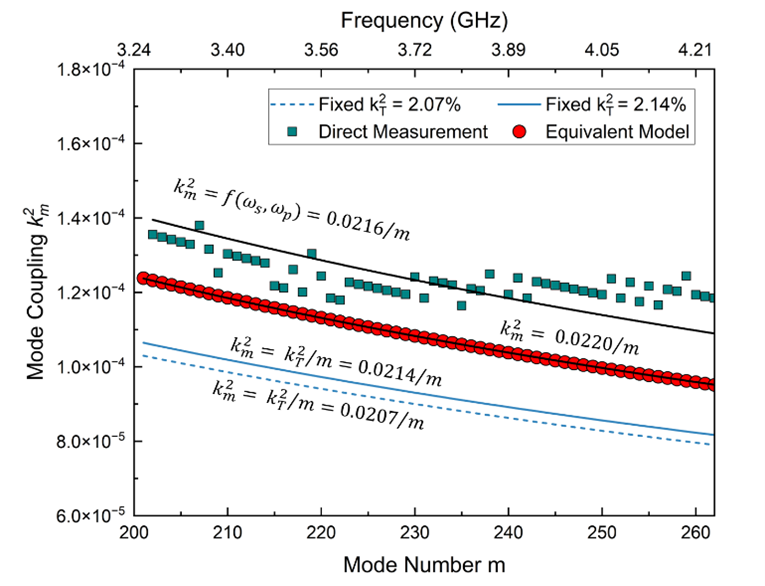}
	\caption[km2]{Trends showing measured and modeled values of the effective modal coupling coefficients as a function of the piezoelectric coupling coefficient of the transducer (GaN). The measured values are limited by equipment resolution and noise. Measured and modeled values clearly follow a  $ k_m^2=(k_T^2)⁄m $  trend, with values of $k_T^2$ close to expected values for GaN. }
	\label{fig:fig5}
\end{figure}

\begin{figure}
	\centering
	\includegraphics[width=0.7\linewidth]{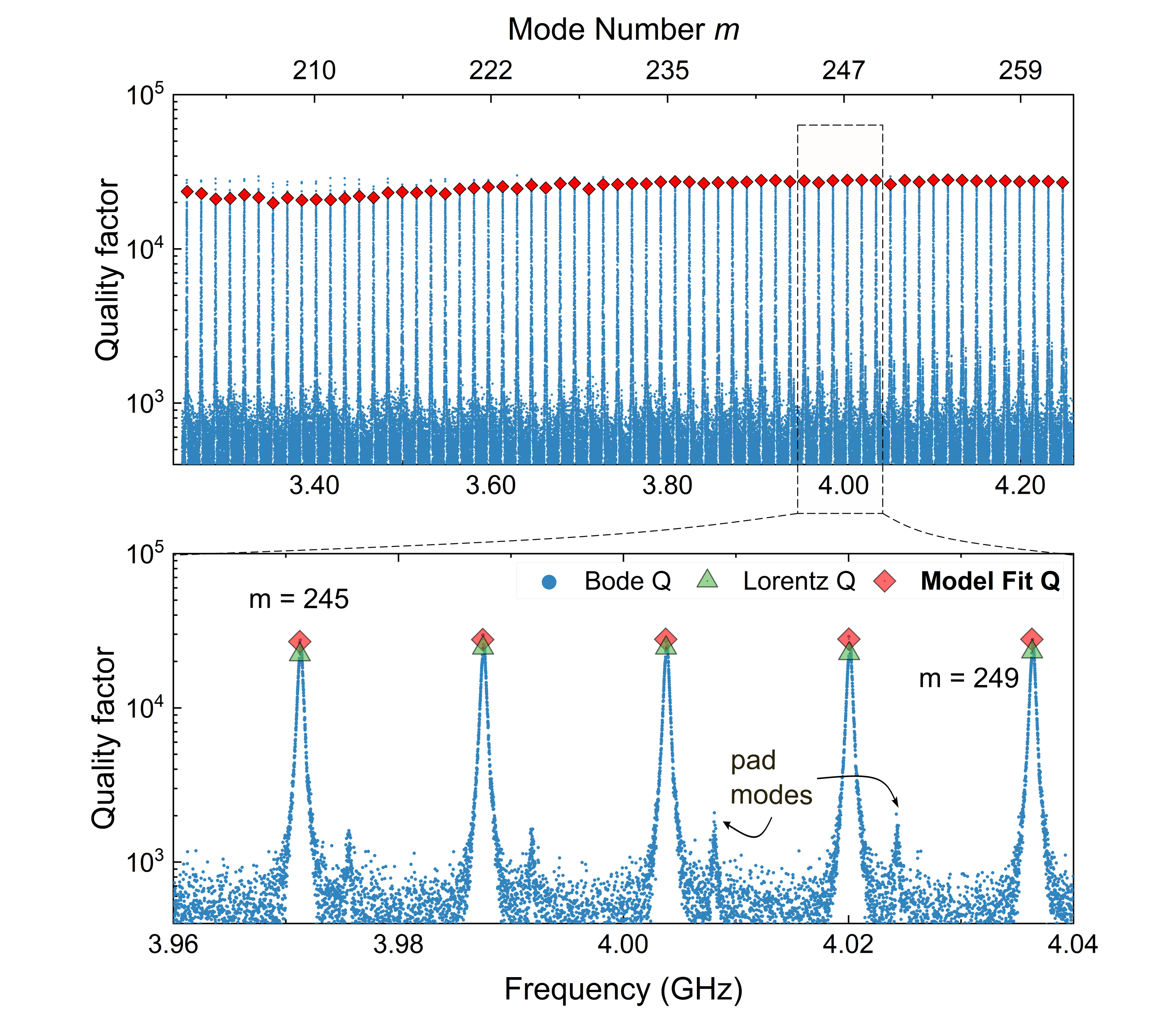}
	\caption[quality factors]{Modeled values of $Q_m^{'}$ agree very closely with the \(Q_{Bode}\) values directly extracted from the measured data, as well as the \(Q_{Lor}\) values derived by fitting individual modes from the dataset to a Lorentzian function. The close match between measured and modeled values validate the model, data-fitting, and parameter extraction approach. For the GaN/NbN/Sapphire measured at room temperature, we observe Q values on the order of $3 \times 10^4$ and $f_m \times Q_m$ values on the order of \(10^{14}\) Hz.}
	\label{fig:fig6}
\end{figure}

The modeled values of the quality factor $Q_m^{'}$ are calculated using (\ref{eq22}), and compared with both the directly extracted \(Q_{Bode}\) values \cite{RN1837}, and (for a representative subset of modes) the \(Q_{Lor}\) values from fitting the data to Lorentzian functions. The close matching between all three methods indicates good convergence for the model fitting (Fig. \ref{fig:fig6}). The values of  $Q_m^{'}$ for this GaN/NbN/Sapphire HBAR, measured at room temperature lie in the range of $2 \times 10^4 - 3 \times 10^4 $ across the measured spectral range, with the figure of merit $f×Q$ on the order of $10^{14}$ Hz.

\section{Extended HBAR Equivalent Circuit Model}
\label{sec:SecIV}

\subsection{Higher order transducer envelopes}

In the discussion so far, we have only considered the fundamental mode of the piezoelectric transducer, represented by \textbf{\textit{T}}. Like all resonators, the piezoelectric transducer itself has overtones. Assuming the free piezoelectric transducer is symmetric about the horizontal axis, we should expect to see only odd transducer harmonics (i.e., the $1^{st}$, $3^{rd}$, $5^{th}$, …). This symmetry can be slightly disturbed for electrodes with unequal thicknesses, or using different materials, or it can be completely broken for composite piezoelectric transducers, leading to the possibility of even harmonics \cite{RN2732,RN378,RN2597}. For higher modes $n$ of the piezoelectric transducer, we expect to see transducer coupling efficiency scale down drastically \cite{RN3020,RN1711}:

\begin{equation} 	k_{T_n}^2  \approx \frac{k_{T_1}^2}{n^2}  \label{eq25} \end{equation}

\begin{figure*}
	\centering
	\includegraphics[width=1\linewidth]{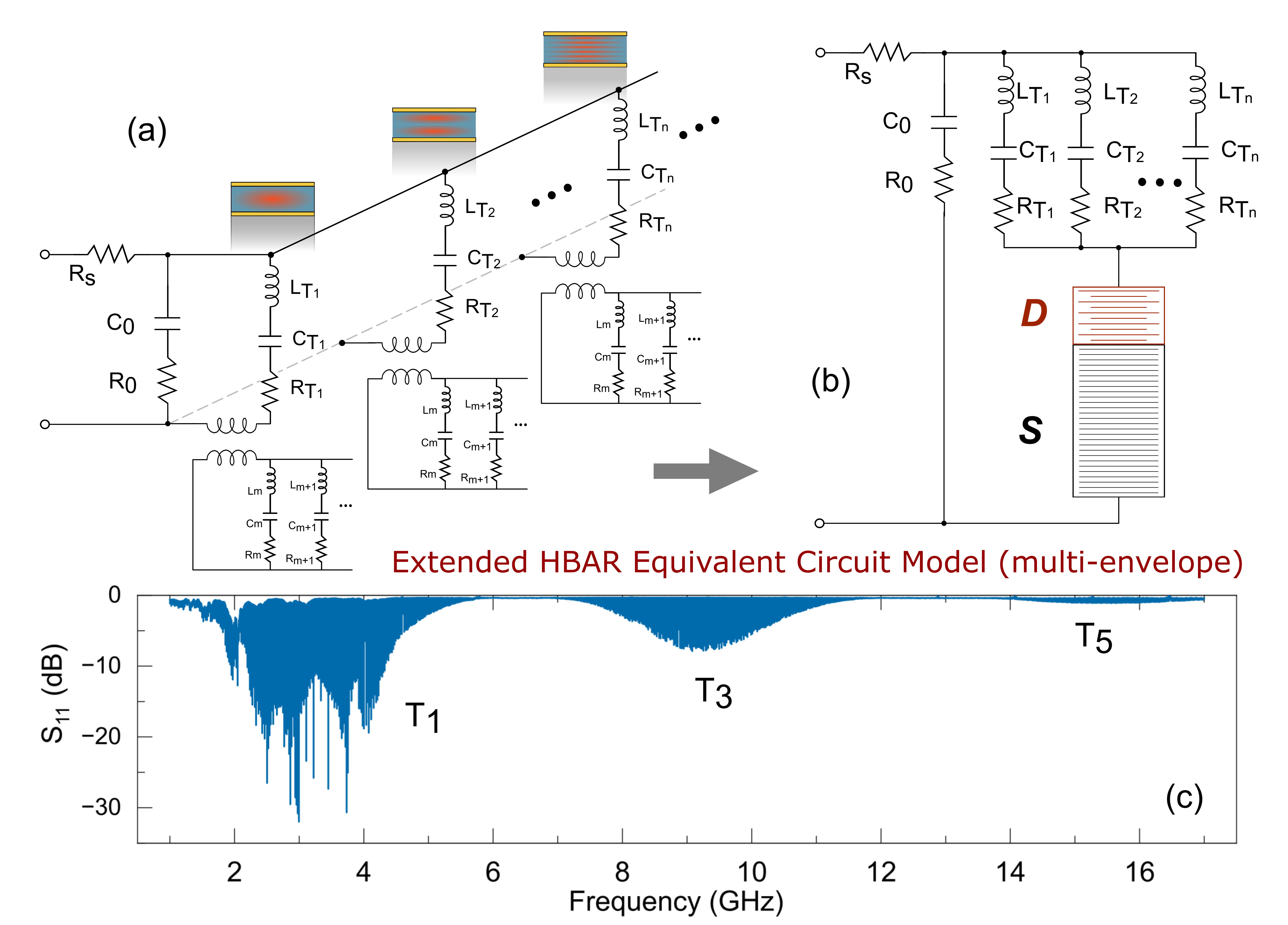}
	\caption[extendedhbarmodel]{The extended HBAR equivalent circuit model. (a) A single transducer can operate at any of its $n$ overtones. A significant advantage of the new HBAR equivalent circuit model is that it can be extended to cover all overtones of the transducer by (b) introducing multiple transducer branches $\textbf{\textit{T}}_n$ linked to the same detuning coupler and substrate elements $\textbf{\textit{D}}$ and $\textbf{\textit{S}}$. (c) A representative HBAR spectrum showing three transducer overtones and envelopes for a GaN/NbN/SiC HBAR. Here, $\textbf{\textit{T}}_n$ are odd harmonics of the transducer such that envelopes are centered at $\sim$3 GHz, $\sim$9 GHz and $ \sim$15 GHz. Data are adapted from previously published work \cite{RN2249}.}
	\label{fig:fig7}
\end{figure*}

At the cost of lower piezoelectric coupling efficiency, higher transducer modes can be used to increase the operating range of the HBARs and have been used to generate higher order transducer envelopes, each of which contains multiple HBAR modes. A significant advantage of working with higher envelope modes at higher frequencies is that the HBAR cavity operates in the higher frequency Landau-Rumer regime where anharmonic phonon loss scales more favorably with both frequency and cooling as compared the Akhieser regime \cite{RN2397}. For applications such as quantum acoustics, where the resonator can be cooled to very low temperatures, one could use higher transducer envelopes to generate operating in the Landau-Rumer regime  with a high figure of merit $f_m \times Q_m$.	In an extension of the HBAR equivalent circuit (Fig.  7), each $n^{th}$  mode of the transducer can now be modeled as a separate transducer branch represented by $\textbf{\textit{T}}_n=\{ L_{T_n},C_{T_n},R_{T_n} \}$ such that 

\begin{equation}	\omega_{T_n} = \frac{1}{\sqrt{L_{T_n} C_{T_n}}}		\label{eq26} \end{equation}

\begin{equation} k_{T_n}^2 = \frac{\pi}{8} \frac{C_{T_n}}{C_0} \left( \frac{C_0-C_{T_n}}{C_0} \right) \approx \frac{k_{T_1}^2}{n^2}  \label{eq27} \end{equation}

Each such transducer branch generates an envelope of HBAR modes centered around $\omega_{T_n}$. These HBAR modes are completely described by combining $\textbf{\textit{T}}_n$ with the same substrate and interface elements \textbf{\textit{S}} and \textbf{\textit{D}} since the perfectly periodic modes of the bare substrate and the aperiodicity values  $a_m $ are already encapsulated therein for all integer values of $m$.

\subsection{Multiple Transducers on the same substrate}

\begin{figure}
	\centering
	\includegraphics[width=0.7\linewidth]{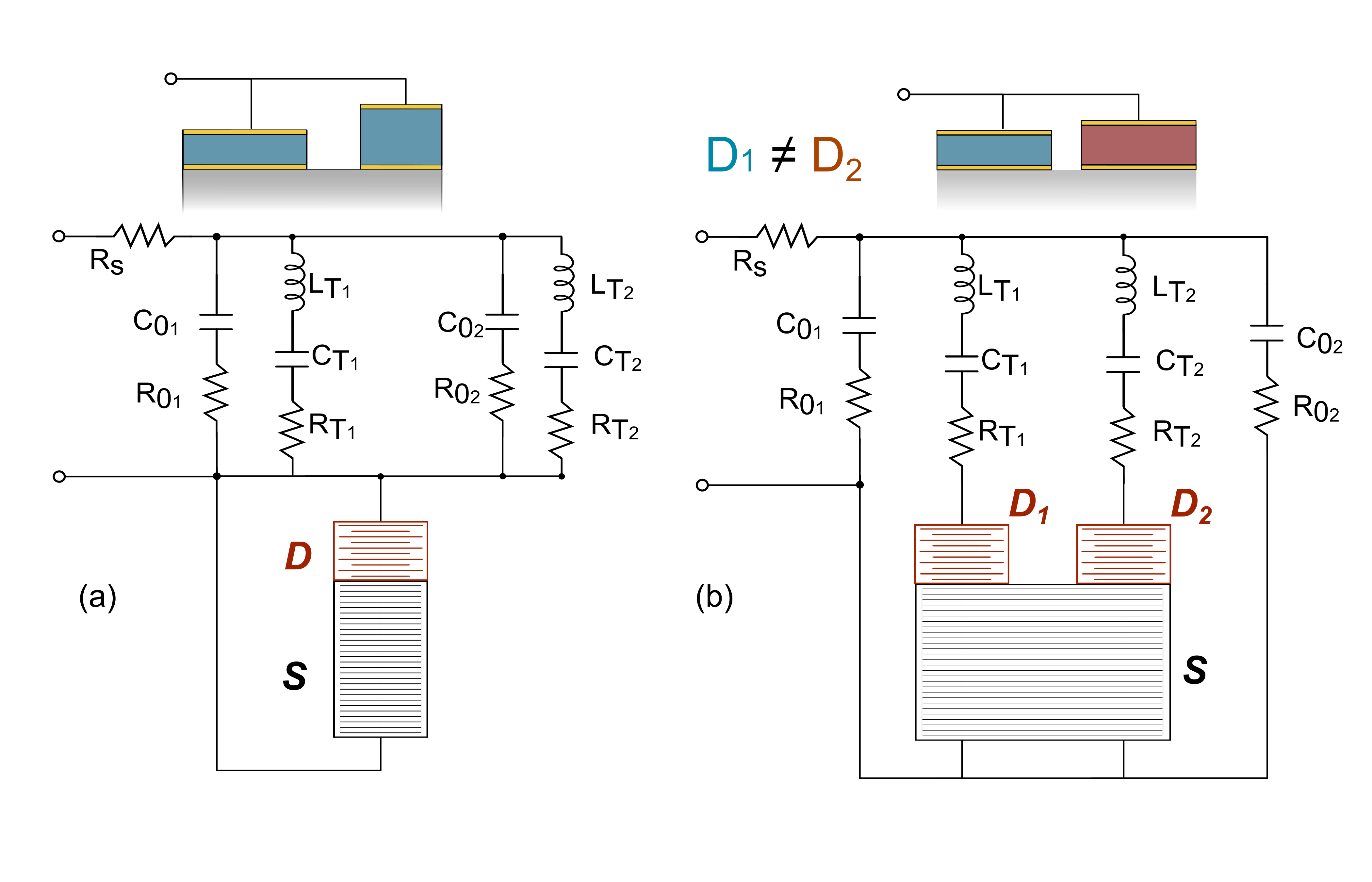}
	\caption[multipletransducersv2]{The new and compact HBAR equivalent model can be used to  (a) model multiple transducers of the same material on the same substrate and (b) model multiple transducers with dissimilar materials on the same substrate in a compact manner.}
	\label{fig:fig8}
\end{figure}

The model for higher order transducer modes $\omega_{\textbf{\textit{T}}_n}$  can be generalized to account for multiple transducers on the same substrate, linked electrically. Multiple transducers with different thickness or area could be fabricated on the same substrate and linked in parallel electrically to cover a broader envelope, or provide some unique functionality as compared to a single transducer (Fig. \ref{fig:fig8}). In such a generalized configuration, the extended model equivalent circuit from Fig. \ref{fig:fig7} can be used, with the modification that $n$ would simply represent the different transducers. A special case in this context deals with the damped ‘pad modes’ that are commonly observed when a simple, single-transducer HBAR is probed by placing thick bond pads, wire bonds, or probes on part of the top electrode. While pad modes can be avoided at the cost of a more complicated fabrication process, the addition of a second, highly damped, transducer branch can model these pad modes. 

Note that multiple transducers made with different materials (for either the piezoelectric or the electrodes, especially the bottom electrode) do not fit into the model shown in Fig.\ref{fig:fig8}(a). In such a scenario, even though the substrate is identical, electrical capacitance and the acoustic impedance between the substrate and different transducers are not the same, i.e., $a_{m_1}  \neq a_{m_2}$, and $D_1  \neq D_2$. Such a configuration could be modeled as shown in Fig.\ref{fig:fig8}(b). In these scenarios involving multiple transducers, $n$ is simply the number of transducers, and there is no required internal harmonic relationship between $\omega_{\textbf{\textit{T}}_n}$ or the corresponding envelopes. If $\omega_{\textbf{\textit{T}}_n}$ are not well separated in frequency, the transducer envelopes could be close to each other or even overlap. These transducer modes can no longer be considered fully independent branches and will present a harder problem for model fitting and parameter extraction.

\subsection{Acoustically coupled HBAR filters}
Acoustically coupled filters often use a modified version of the mBVD model to represent input and output resonators and the acoustic coupling between them \cite{RN2018,RN2011}. Acoustically coupled HBAR filters with a comb-like filter spectrum have been demonstrated experimentally \cite{RN1830,RN2346,RN2399}. The HBAR equivalent circuit framework detailed here could be expanded to describe the coupled modes of the HBAR filter in the future; this is beyond the scope of the current work.

\section{Modeling Perturbations to the HBAR}
\label{sec:SecV}

The new equivalent circuit model can be used to simplify analysis of perturbations to the nominal state of the HBAR. HBARs and HBAR-based oscillators have been used as classical sensors for acceleration sensing, gravimetry, mass-loading, pressure, fluid flow and viscosity, and temperature \cite{RN3033,RN2009,RN2335,RN3031}. Additionally, HBAR cavity phonons can interact with superconducting and spin qubits, and have been proposed as detectors for a variety of less explored physical phenomena such as gravitational waves and dark matter which can directly or indirectly couple to GHz-frequency acoustic phonons \cite{RN2892}. For all these cases, while the underlying interactions between the perturbation and the HBAR are complex and require detailed study and full analytical modeling, the HBAR equivalent circuit can be a useful technique to model the behavior of the system phenomenologically. 

The advantage of using the new equivalent model is that it allows us to independently model perturbations and coupling mechanisms to separate sub-components of the HBAR; either the transducer, the cavity, or both. Keeping the sub-components separate also allows us to model any non-linearities faithfully; this would be harder to do using (say) the multi-mode mBVD model, where the entire resonance mode is lumped into a single branch.

\begin{figure}
	\centering
	\includegraphics[width=0.7\linewidth]{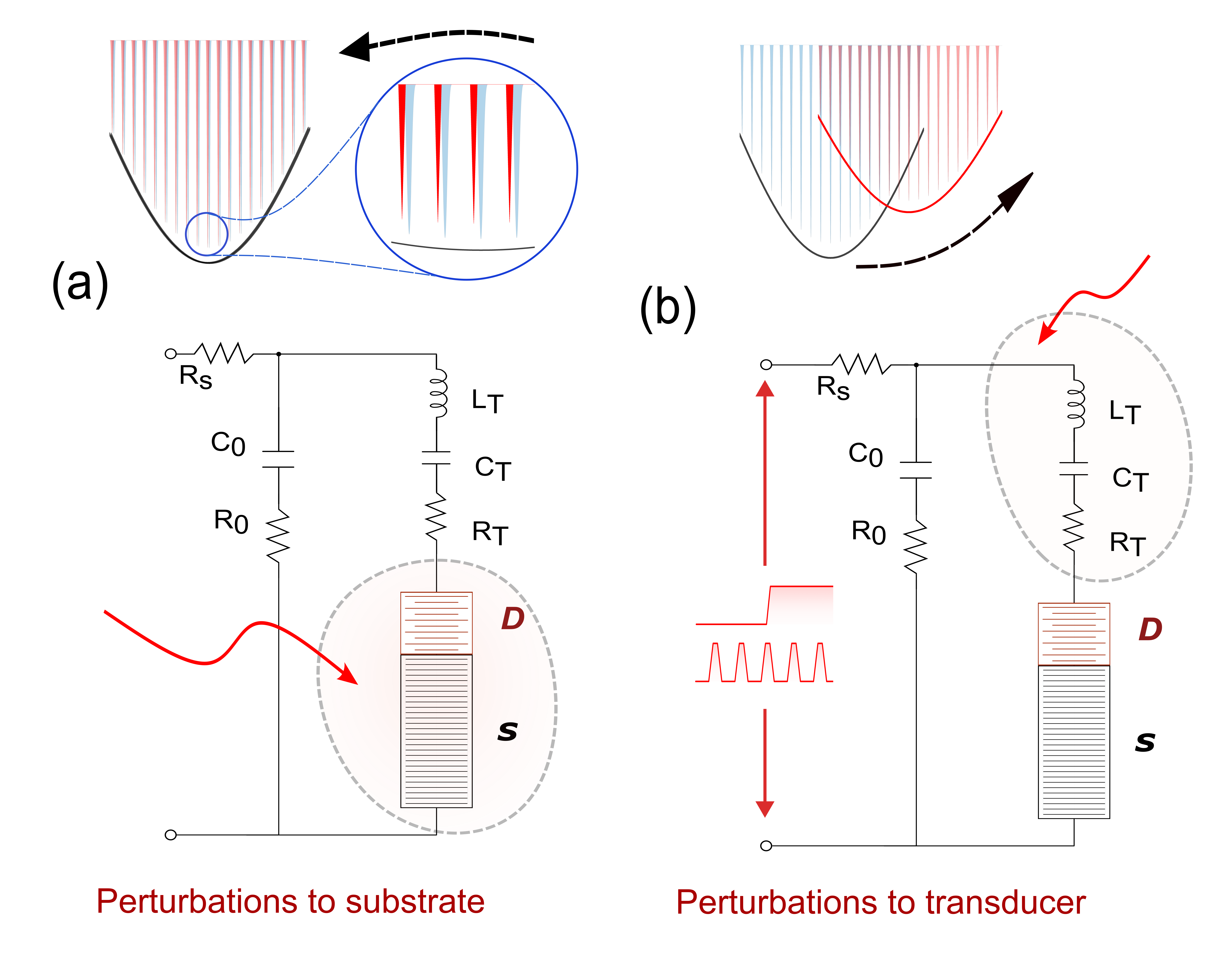}
	\caption[perturbations]{Perturbations to the HBAR can be modeled as perturbations (a) to the substrate alone, resulting in a small shift in all the cavity phonon modes, or (b) to the the transducer alone, which could result in large changes to the envelope while keeping the the cavity phonon frequencies unchanged, or (not shown) to both substrate and transducer at the same time.  Purely dispersive perturbation would change the the non-dissipative elements \textit{L} and \textit{C}, while dissipative perturbations would impact \textit{R}. In practical scenarios, we often encounter a combination of dispersive and dissipative effects.}
	\label{fig:fig9}
\end{figure}

Purely dispersive coupling mechanisms would result in modulation of the non-dissipative elements $L_T$,  $C_T$, $L_m$, and $C_m$. Purely dissipative mechanisms would affect the elements $R_T$ and $R_m$. Practically, many perturbations to the HBAR would need to be modeled as a combination of dispersive and dissipative coupling. Here, we do not go into detailed models for any specific scenario but simply mention some examples to highlight the general utility of the new HBAR equivalent circuit model.

\subsection{Perturbations to the entire HBAR heterostructure}

Ambient and inertial frame changes (e.g., temperature, hydrostatic pressure, acceleration, rotation) will affect both the transducer and the substrate, albeit at different rates depending on the materials involved. In many such cases, the large disparity in thickness between the substrate and the transducer could allow us to ignore the transducer. However, if needed, we can model common perturbations by allowing changes to all relevant virtual elements in the HBAR equivalent circuit. Such a ‘full-model’ approach is expected to be critical for performing multi-parameter experiments to separate phonon loss mechanisms.   

\subsection{Perturbations to the substrate alone}

The authors recently demonstrated HBARs on ferrimagnetic yttrium iron garnet (YIG) substrates \cite{RN2610}. At appropriate magnetic fields, the HBAR phonons hybridize with magnons in the YIG substrate. The broadband magnetoelastic hybridization impacts a range of HBAR modes simultaneously causing mode tuning (dispersive coupling) as well as mode suppression (dissipative coupling). The piezoelectric transducer is unaffected by the magnetic field. This magnetoelastic HBAR can now be simply modeled by using an HBAR substrate element $\textbf{\textit{S}}(B_M)$  that is dependent on the magnetic field $B_M$. A similar approach could be taken to model the strain coupling between one or more HBAR mode and magnetically sensitive spin defect ensembles within the substrate \cite{RN1959}. 

\subsection{Perturbations to the piezoelectric transducer alone}

The simplest way to interact with the piezoelectric transducer alone is via an electric field applied across it. Dynamic DC bias tuning of FBARs is a well-established technique often used to tune mode frequencies over a short range, or to compensate for environmental effects. For linear electrostrictive materials (including AlN, GaN, and ScAlN) the effects are small, often resulting in frequency shifts on the order of $\sim \pm$10 ppm/V \cite{RN3052}. More significant changes to frequency and coupling coefficients are observed by using ferroelectric materials as the transducer: the transducer can be tuned to over a larger range, completely switched off, or switched between harmonics by the application of a DC bias \cite{RN3053}. 

In all of these scenarios, the electric field is applied between the electrodes of the transducer. We can model the tuning as a change to $L_T$ and $C_T$ for purely dispersive effects, and to $R_T$ for any attendant dissipation. For ferroelectric materials, we should also consider any changes in $\textbf{\textit{E}}$ as a function of applied bias. However, any such bias tuning should have no effect on \textbf{\textit{S}} or \textbf{\textit{D}}, since the field is not applied across the substrate. Experimental demonstration of this effect in $Ba_{0.5}Sr_{0.5}TiO_{3}$-on-sapphire ferroelectric HBARs is provided by Sandeep \textit{et al} \cite{RN3034,RN2985}, who demonstrate large bias tuning of the transducer envelope frequency and the piezoelectric coupling coefficient $k_T^2$ (transducer attributes) without any change to the mode spacing (substrate attributes).

An important contemporary and relevant scenario is quantum acoustics, with applications involving coupling of superconducting qubits to the high-\textit{Q} phonon modes of the HBAR. This coupling models the effect of qubits’ charge or electric field on the piezoelectric transducer \cite{RN1827,RN1829}. The coupled system consisting of the frequency-tunable qubit and HBAR phonons can now be used to perform spectroscopy on the qubit, create non-classical mechanical states, and act as readout for quantum metrology. Superconducting circuits are electrical systems and are often described by lumped element networks comprised of voltage/current/charge sources, passive electrical elements, electromagnetic ($LC$) resonators, and the non-linear elements such as Josephson junctions that make up the qubit. While detailed full-physics descriptions of coupled qubit-HBAR experiments are derived in the literature (albeit often for single or just a few HBAR modes), the use of this HBAR equivalent model gives us a powerful technique to incorporate and include the full HBAR and its complete phonon spectrum in the same lumped element superconducting circuit network as the electromagnetic elements \cite{RN1827,RN1829}.

\section{Model Limitations and Caveats}

It is important to reiterate that the HBAR model, like all equivalent circuit models, is an approximation. There are caveats to its use that must be considered. First, it is possible that there are multiple possible solutions that fit any measured dataset; only one will correspond to physical reality. While computationally more intensive, measuring many modes across a wide range of frequencies (wavelengths) could reduce fitting error and thus give us more accurate insights into the physics of the HBAR.
 
Secondly, model fitting and parameter extraction works best with sharp high-$Q$, well separated HBAR modes. Analogous to photonic cavities, we can define the acoustic finesse $\textbf{\textit{F}}$ of the HBAR as the ratio of mode spacing and linewidth (full width at half maximum or $FWHM$)

\begin{equation} 	\textbf{\textit{F}} = \frac{FSR}{FWHM} = \frac{Q_m^{'} \times FSR }{\omega_m^{'}/2 \pi} \label{eq28} \end{equation}

A high acoustic finesse implies narrow spectral confinement of phonon energy with respect to the FSR. If the HBAR is used as a sensor, high finesse also indicates good sensing resolution. For clean parameter extraction, it is important to work with high finesse HBARs $\textbf{\textit{F}} \gg 10$. For low finesse HBARs, the parameters of any branch will be dependent on the adjacent branches, complicating the model fitting and extraction. For low finesse HBARs, especially important when working with new materials or at very high frequencies, more sophisticated computational techniques could be used to improve accuracy in the future. Note that the measured finesse is $ \textbf{\textit{F}} \sim 120 $ for the HBAR dataset discussed in Section \ref{sec:SecIII}.

Finally, while we have shown that the analogy between damped harmonic oscillators in the electrical and mechanical domains hold true, interpreting numerical values of the equivalent circuit is harder. There are some uncertainties in the model; we can describe any mode by using the product of $L_m^{'}$ and  $C_m^{'}$  but it is harder to separate $L_m^{'}$ and  $C_m^{'}$ and assign unambiguous physical meaning to them in terms of the equivalent modal mass and modal stiffness $M_m$ and $K_m$. As described earlier, there is an uncertainty in the dissipative chain $R_m^{'}$ . While the total loss can be modeled accurately, it is harder to separate the contributing loss mechanisms. 

\section{Conclusion}

The new HBAR equivalent circuit developed here models the rich, dense, and broadband dynamics of HBARs while retaining the physical significance of its constituent elements and maintaining internal physical relationships. This is an intuitive model that can be easily interpreted or manipulated by scientists and engineers with a basic grasp of harmonic oscillators, and circuit theory. The new model is also scalable; a variety of configurations and perturbations to the nominal state can be easily and completely described. 
In addition, the new model can form the kernel for fitting and analyzing large, measured datasets. Computationally, while it has more elements than the multi-mode mBVD model, many of the new elements are strictly related to the mode number and some constant values, reducing complexity by using a smaller number of tightly constrained independent variables for fitting. Future work shall include the use of this model in conjunction with deep learning techniques for efficient data fitting, parameter extraction and analysis of large datasets.

\section*{Acknowledgments}
The authors are grateful to Dr. D. Scott Katzer (NRL) for providing the GaN/NbN/sapphire HBAR heterostructure used in this work. The authors also thank Dr. Guilhem Ribeill (RTX BBN), Dr. Gerasimos Angelatos (RTX BBN), Dr. James Champlain (NRL), and Dr. George Stantchev (NRL) for many useful discussions that led to the development and refinement of this model.

\end{document}